\documentclass[useAMS,usenatbib]{mn2e}
\usepackage{graphicx,times}
\usepackage{bm}
\usepackage{epsf}
\usepackage{amsmath}
\usepackage{amssymb}
\usepackage{graphics}
\usepackage{epstopdf}
\usepackage{float}
\usepackage{url}
\usepackage{color}

\begin{document}

\onecolumn
\title[Modification of the halo  mass function by kurtosis associated with primordial non-Gaussianity]
{Modification of the halo  mass function by kurtosis associated with primordial non-Gaussianity}

\author[S. Yokoyama, N. Sugiyama, S. Zaroubi and J. Silk]
{Shuichiro Yokoyama$^{1}$, Naoshi Sugiyama$^{1,2}$, Saleem Zaroubi$^{3,4}$ and Joseph Silk$^{5}$ \\
$^{1}$
Department of Physics and Astrophysics, Nagoya University, Aichi 464-8602, Japan\\
$^{2}$
Institute for the Physics and Mathematics of the Universe,
University of Tokyo, Kashiwa, Chiba, 277-8568, Japan\\
$^{3}$
Kapteyn Astronomical Institute,
University of Groningen,
P.O. Box 800, 9700 AV Groningen, the Netherlands\\
$^{4}$
Physics Department, Technion, Haifa 32000, Israel\\
$^{5}$
Oxford University, Astrophysics, Denys Wilkinson Building,
Keble Road, Oxford, OX1, 3RH, UK
}
\maketitle

\begin{abstract}
We study
the halo mass function in the presence of a  kurtosis type of primordial non-Gaussianity.
The kurtosis corresponds to the trispectrum as defined in Fourier space.
The primordial trispectrum is commonly characterized by two parameters, 
$\tau_{\rm NL}$ and $g_{\rm NL}$.
We focus on $\tau_{\rm NL}$ which is an important parameter to test
the physics of multi-field inflation models.
As applications of the derived non-Gaussian mass function,
we consider the effects on the abundance of void structure, 
on early star formation, and on formation of the most massive objects at high redshift.
We show that by comparing the effects of primordial non-Gaussianity on
cluster abundance with that on void abundance,
we can distinguish between the skewness and the kurtosis types 
of primordial non-Gaussianity.
As for early star formation,
we show that the kurtosis type of primordial non-Gaussianity 
seems on the average not to affect the reionization history of the Universe.
However, at high redshifts  (up to $z\simeq 20$)
such non-Gaussianity does somewhat affect the early stages of reionization.
\end{abstract}


\section{Introduction}

The inflation paradigm
has been well-known
as a successful scenario
for resolving several shortcomings of the standard
Big Bang Model, in particular,
the generation of primordial fluctuations
which seed cosmic microwave background (CMB)
fluctuations
and structure formation
of the Universe.
In the standard inflationary scenario, 
the primordial density fluctuations
are generated from quantum fluctuations of a
scalar field
and they have almost Gaussian statistics.
In recent years
it has been realized that studying the non-Gaussianity
of the primordial density fluctuations can reveal
valuable information about the dynamics of inflation
\citep{Komatsu:2001rj,Bartolo:2004if,Bartolo:2010qu,Komatsu:2010hc} (and references therein).
Thanks to  significant progress in cosmological
observations, most notably the CMB observations,
we may expect that
a meaningful measurement of this quantity
will become observationally available in the
near future and will thereby allow
several inflation models to be  tested.

In Ref.~\citep{Komatsu:2001rj}, the authors have introduced a
simple new parameter
which describes the deviation from Gaussianity of the statistics of the
primordial curvature fluctuations, the so-called non-linearity parameter
$f_{\rm NL}$, defined as~\citep{Salopek:1990jq,Gangui:1993tt,Verde:1999ij}
\begin{eqnarray}
\zeta({\bf x}) = \zeta_{\rm G}({\bf x}) 
+ {3 \over 5}f_{\rm NL}\left( \zeta_{\rm G}^2({\bf x})
- \langle \zeta_{\rm G}({\bf x})^2 \rangle \right)
+ O(\zeta_{\rm G}^3({\bf x}))~,
\label{eq:originalfNL}
\end{eqnarray}
where $\zeta$ represents the primordial curvature fluctuations on a  uniform energy density hypersurface
and $\zeta_{\rm G}$ denotes the
Gaussian part.
In the Probability Density Function (PDF) of the primordial fluctuations,
the non-zero value of the non-linearity parameter $f_{\rm NL}$
may generate a non-zero value of the skewness (3rd order moment), the kurtosis(4-th order moment)
and so on.
Obviously,
the skewness can be parametrized by the leading term using $f_{\rm NL}$.
However, the kurtosis can be 
affected not only by the $f_{\rm NL}$ term but also
by higher order terms, such as the $\zeta_{\rm G}^3({\bf x})$ term in the above expression
(\ref{eq:originalfNL}).
In general, one needs two parameters in order to characterize the kurtosis
in the PDF. These parameters are normally
called $\tau_{\rm NL}$
and $g_{\rm NL}$,
where the first is usually (although not always) related to $f_{\rm NL}$
and the second is the parameter that
characterizes the third moment of $\zeta$.
Such kind of non-linearity is the so-called local type of non-Gaussianity.
Recently, other types of non-Gaussianity have been discussed in the literature,
\textit{e.g.}, equilateral and orthogonal types.
Theoretically, the local type of non-Gaussianity
can be generated from the super-horizon non-linear dynamics of 
primordial curvature perturbations.
On the other hand,
the equilateral and orthogonal types of non-Gaussianity
can be generated when one considers a 
scalar field which has a 
non-canonical kinetic term or the
higher order derivative correction terms.
In this paper, we focus on the local type of non-Gaussianity
and consider the case where the equilateral and orthogonal types are negligible.

In the case where the primordial curvature fluctuations
were generated from single field stochastic fluctuations (single-sourced case), \textit{i.e.},
the primordial curvature fluctuations can be expressed as Eq.~(\ref{eq:originalfNL}),
and $\tau_{\rm NL}$ can be described only by $f_{\rm NL}$.
But in general, \textit{e.g.}, if the primordial curvature fluctuations
were generated from multi-stochastic fluctuations then
$\tau_{\rm NL}$ and $f_{\rm NL}$ have no universal relation any more~\citep{Suyama:2007bg,Suyama:2010uj,Sugiyama:2011jt}.
Hence, it seems to be important to investigate the observational consequences of $\tau_{\rm NL}$
independently of $f_{\rm NL}$.

In this paper, we focus on the effects of the kurtosis type of primordial non-Gaussianity on
the Large Scale Structure (LSS), in particular, on the halo mass function.
There are
many studies
of the effects of primordial non-Gaussianity on
the LSS and also on the formulation of the non-Gaussian halo mass function
\citep{Matarrese:2000iz,Slosar:2008hx,Maggiore:2009rx,
Verde:2010wp,D'Amico:2010ta,DeSimone:2010mu,Wagner:2010me} (and references therein),
which focus  not only on
$f_{\rm NL}$-type but also $g_{\rm NL}$-type~\citep{Desjacques:2009jb,Maggiore:2009hp,Chongchitnan:2010xz,Chongchitnan:2010hb,Enqvist:2010bg}.
Here, we study the effects of kurtosis of the non-Gaussian primordial fluctuations
whose non-linearity is parameterized by the two free parameters, $g_{\rm NL}$ and $\tau_{\rm NL}$.
Recently, 
a number of authors have studied
non-Gaussian initial perturbations in two-field inflationary models
\citep{Tseliakhovich:2010kf,Smith:2010gx}.
In these papers, the authors have considered the effect
of non-Gaussianity
on the halo bias.
Although this type of primordial non-Gaussianity is similar to the one considered here,
we study the effects on the halo mass function.

This paper is organized as follows.
In the next section,
we 
briefly review 
the kurtosis type of 
primordial non-Gaussianity 
considered here.
In section~\ref{sec:nGmass},
we formulate halo mass functions 
with  primordial non-Gaussianity,
based on the Press-Schechter theory
and Edgeworth expansion.
In section~\ref{sec:nGappl},
we apply the non-Gaussian halo mass function to the formation of astrophysical objects.
We consider three applications:
early star formation, the most massive object at high redshift
and
the abundance of voids.
Section~\ref{sec:sum} provides a discussion and summary of our results.
We adopt throughout the best fit cosmological parameters taken from WMAP 7-year data.

\section{Trispectrum of primordial non-Gaussian curvature fluctuations}

Here, we focus on the local-type non-Gaussianity.
Following the notation commonly used,
in the single-sourced case, up to the third order,
the primordial curvature fluctuations can be expressed as
\begin{eqnarray}
\zeta = \zeta_{\rm G} + {3 \over 5}f_{\rm NL}
\left( \zeta_{\rm G}^2 - \langle \zeta_{\rm G}^2 \rangle \right)
+ {9 \over 25} g_{\rm NL} \zeta_{\rm G}^3~.
\label{eq:zetafNL}
\end{eqnarray}
Based on this expression, the trispectrum of $\zeta$
is given by
\begin{eqnarray}
&&\langle
\zeta({\bf k}_1) \zeta({\bf k}_2) \zeta({\bf k}_3) \zeta({\bf k}_4)
\rangle = (2 \pi)^3 T_\zeta(k_1,k_2,k_3,k_4)
\delta^{(3)}\left( {\bf k}_1 + {\bf k}_2 + {\bf k}_3 + {\bf k}_4 \right)~,\cr\cr
&&
T_\zeta(k_1,k_2,k_3,k_4) = \tau_{\rm NL}
\left( 
P_\zeta(k_1)P_\zeta(k_2)P_\zeta(k_{13}) + 11 {\rm perms.}\right)
+ {54 \over 25}g_{\rm NL}
\left( 
P_\zeta(k_1)P_\zeta(k_2)P_\zeta(k_{3}) + 3 {\rm perms.}\right)~,
\label{eq:trispectrum}
\end{eqnarray}
where 
$k_{13}=\left| {\bf k}_1 + {\bf k}_3\right|$ and
$P_\zeta(k_1)$ is a power spectrum of $\zeta$ given by
$\langle \zeta({\bf k}_1) \zeta({\bf k}_2) \rangle = (2\pi)^3 P(k_1)\delta^{(3)}
\left( {\bf k}_1 + {\bf k}_2 \right)$.
For the above definition of $\tau_{\rm NL}$ and the form of the non-linearity of the
curvature perturbation~(\ref{eq:zetafNL}),
$\tau_{\rm NL}$ can be written in terms of the non-linearity parameter $f_{\rm NL}$ as
\begin{eqnarray}
\tau_{\rm NL} = {36 \over 25}f_{\rm NL}^2~.
\end{eqnarray}
This consistency relation is satisfied
only  in the case where the primordial curvature fluctuations
can be described by Eq.~(\ref{eq:zetafNL}), namely,
the primordial curvature fluctuations are sourced only from the
quantum fluctuations of a single scalar field, \textit{e.g.}, curvaton~\citep{Enqvist:2001zp,Lyth:2001nq,Moroi:2001ct}.

However, if there are multiple sources of the primordial curvature fluctuations,
then the above consistency relation is not satisfied~\citep{Langlois:2004nn,Ichikawa:2008iq,Huang:2009vk,Byrnes:2010em}.
In general, it has been known that
there exists an inequality between
the local type non-linearity parameters $\tau_{\rm NL}$ and 
$f_{\rm NL}$ given by~\citep{Suyama:2007bg,Suyama:2010uj,Sugiyama:2011jt}
\begin{eqnarray}
\tau_{\rm NL} > {1 \over 2}\left({6 \over 5}f_{\rm NL}\right)^2~.
\end{eqnarray}
For example,
let us consider the local-type non-Gaussianity given by
\begin{eqnarray}
\zeta = \phi_{\rm G} + {3 \over 5}f_{\rm NL} 
\left( \phi_{\rm G}^2 - \langle \phi_{\rm G}^2 \rangle \right)
+ t_{\rm NL} \phi_{\rm G} \psi_{\rm G}~,
\label{eq:multi}
\end{eqnarray}
where
$\phi_{\rm G}$ and $\psi_{\rm G}$ are Gaussian fluctuations with
$\langle \phi_{\rm G} \psi_{\rm G}\rangle = 0$ and
$t_{\rm NL}$ is a non-linearity parameter, which represents 
the non-linear coupling between $\phi_{\rm G}$ and $\psi_{\rm G}$
in $\zeta_{\rm G}$.
At  leading order,
the power spectrum of $\zeta$ is given by that of the Gaussian part $\phi_{\rm G}$ as
\begin{eqnarray}
\langle \zeta({\bf k}) \zeta({\bf k}') \rangle 
= \langle \phi_{\rm G}({\bf k}) \phi_{\rm G}({\bf k}') \rangle
= (2\pi)^3 P_\phi(k) \delta^{(3)}({\bf k}+ {\bf k}')~,
\label{eq:powerspectrum}
\end{eqnarray}
and the bispectrum is given only by $f_{\rm NL}$ as
\begin{eqnarray}
\langle \zeta({\bf k}_1) \zeta({\bf k}_2) \zeta({\bf k}_3) \rangle 
= (2\pi)^3 {6 \over 5} f_{\rm NL} (P_\phi(k_1)P_\phi(k_2) + 2 {\rm perms.})
\delta^{(3)}({\bf k}_1 + {\bf k}_2 + {\bf k}_3)~,
\label{eq:bispectrum}
\end{eqnarray}
because of $\langle \phi_G \psi_G \rangle = 0$.

In the single-source case which corresponds  to the case of $t_{\rm NL} = 0$,
as  mentioned above, the trispectrum can be also parameterized only by $f_{\rm NL}$.
However, for the above type of curvature fluctuations
the trispectrum is given by
\begin{eqnarray}
T_\zeta(k_1,k_2,k_3,k_4) &=&
\left(\frac{6}{5} f_{\rm NL}\right)^2 (P_\phi(k_1)P_\phi(k_2)P_\phi(k_{13}) + 11 {\rm perms.}) \cr\cr
&&
+ t_{\rm NL}^2\biggl(
P_\phi(k_1)P_\phi(k_2)P_\psi(k_{13}) + P_\phi(k_1)P_\phi(k_2)P_\psi(k_{14})
+ P_\phi(k_1)P_\phi(k_3)P_\psi(k_{12}) \cr\cr
&& \qquad\quad
+ P_\phi(k_1)P_\phi(k_3)P_\psi(k_{14}) 
+ P_\phi(k_1)P_\phi(k_4)P_\psi(k_{12}) + P_\phi(k_1)P_\phi(k_4)P_\psi(k_{13}) \cr\cr
&& \qquad\quad
+ P_\phi(k_2)P_\phi(k_3)P_\psi(k_{12}) + P_\phi(k_2)P_\phi(k_3)P_\psi(k_{24})
+ P_\phi(k_2)P_\phi(k_4)P_\psi(k_{12}) \cr\cr
&& \qquad\quad
+ P_\phi(k_2)P_\phi(k_4)P_\psi(k_{23})
+ P_\phi(k_3)P_\phi(k_4)P_\psi(k_{13}) + P_\phi(k_3)P_\phi(k_4)P_\psi(k_{23}) \biggr)
~,
\end{eqnarray}
where $k_{13} = |{\bf k}_1 + {\bf k}_3|$.
We assume that the power spectra of random Gaussian fields $\phi_G$ and $\psi_G$ 
have only weak scale-dependence, that is, the power spectra are respectively
given by
\begin{eqnarray}
P_\phi(k) \equiv {2\pi^2 \over k^3} A_\phi \left( {k \over k_0} \right)^{n_\phi-1}~,
~P_\psi(k) \equiv {2 \pi^2 \over k^3} A_\psi \left( {k \over k_0} \right)^{n_\psi-1}~,
\end{eqnarray}
where 
$k_0$ is a pivot scale and
$\left| n_\phi - 1 \right| \ll 1$ and $\left| n_\psi - 1 \right| \ll 1$.
In such a case, we can rewrite the power spectrum of $\psi_G$ as
\begin{eqnarray}
P_\psi(k) \simeq \alpha P_\phi(k)~,~ \alpha \equiv A_\psi / A_\phi~,
\end{eqnarray}
and then using the ratio of the amplitudes $\alpha$,
the expression for the trispectrum can be reduced to
\begin{eqnarray}
\langle \zeta({\bf k}_1) \zeta({\bf k}_2) \zeta({\bf k}_3) \zeta({\bf k}_4) \rangle 
& \simeq & (2\pi)^3 
\delta^{(3)}({\bf k}_1 + {\bf k}_2 + {\bf k}_3+{\bf k}_4)\left(
{36 \over 25} f_{\rm NL}^2 + \alpha t_{\rm NL}^2 \right)(P_\phi(k_1)P_\phi(k_2)P_\phi(k_{13}) + 11 {\rm perms.})
~. \nonumber\\
\end{eqnarray}
From the above equation and Eq.~(\ref{eq:trispectrum}), we  easily find that
the non-linearity parameter $\tau_{\rm NL}$ is 
\begin{eqnarray}
\tau_{\rm NL} = {36 \over 25} f_{\rm NL}^2 + \alpha t_{\rm NL}^2 \ge  {36 \over 25} f_{\rm NL}^2~.
\end{eqnarray}
Hence,
in the following discussion,
we consider $\tau_{\rm NL}$ independently of $f_{\rm NL}$.

\section{Non-Gaussian mass function induced from primordial non-Gaussianity}
\label{sec:nGmass}
In the previous section,
we have shown that there is a strong theoretical motivation for  considering $\tau_{\rm NL}$
to be independent of $f_{\rm NL}$.
The parameter $\tau_{\rm NL}$ characterizes
the amplitude  of the trispectrum of primordial curvature fluctuations as well as $g_{\rm NL}$.
Here, we briefly review the formula for the halo mass function with not only the non-zero primordial bispectrum
but also the non-zero primordial trispectrum, based on Press-Schechter theory.  

\subsection{Probability Density Function of the smoothed density field with primordial non-Gaussianity}

The matter density linear fluctuations in Fourier space at redshift $z$,
$\delta ({\bf k}, z)$, are given by
the primordial curvature perturbation on a uniform energy density hypersurface $\zeta({\bf k})$ as
\begin{eqnarray}
&&\delta ({\bf k},z)={\cal M}(k)D(z)\zeta({\bf k})~,\\
&& {\cal M}(k) = {2 \over 5}{1 \over \Omega_{m0}}{k^2 \over H_0^2} T(k)~,
\end{eqnarray}
where $\Omega_{m0}$ is the  present density parameter for total non-relativistic matter,
$H_0$ is the Hubble constant, $D(z)$ is a linear growth function and $T(k)$ is a transfer function.
Using these expressions, we can obtain the linear matter power spectrum as
\begin{eqnarray}
&& \langle \delta ({\bf k},z) \delta ({\bf k}', z)\rangle \equiv (2\pi)^3 P_\delta (k,z) \delta^{(3)}({\bf k} + {\bf k}')~,\\
&& P_\delta (k,z) = {2 \pi^2 \over k^3} {\cal M}(k)^2 D(z)^2 {\cal P}_\phi (k)~,
\end{eqnarray}
where ${\cal P}_\phi (k) = k^3 P_\phi(k)/ ( 2 \pi^2 )$.
Following the standard procedure, let us define the smoothed density fluctuation on a given length scale, $R$, as
\begin{eqnarray}
\delta_R = \int {d^3 {\bf k} \over \left(2\pi\right)^3}W_R(k) \delta ({\bf k},z)~,
\end{eqnarray}
where $W_R(k)$ is the Fourier transform of a spherical top-hat window function given by
\begin{eqnarray}
W_R(k) = 3\left({\sin(kR) \over k^3 R^3} - {\cos(kR) \over k^2 R^2}\right)~.
\end{eqnarray} 

In order to take into account primordial non-Gaussianity in the smoothed density fluctuations,
let us consider the PDF of $\delta_R$, $F(\delta_R) d\delta_R$.
The $n$-th central moment for $F(\delta_R)d \delta_R$ is defined as
\begin{eqnarray}
\langle \delta_R^n\rangle \equiv \int^{\infty}_{-\infty}
\delta_R^n F(\delta_R)d\delta_R~,
\end{eqnarray}
and each reduced $p$-th cumulant can be defined as
\begin{eqnarray}
S_p(R) \equiv \frac{\langle \delta_R^p \rangle_c}{\langle \delta_R^2 \rangle_c^{p-1}}~,
\label{eq:cum}
\end{eqnarray}
where a subscript $c$ denotes the connected part of $p$-point function given by
\begin{eqnarray}
&&\langle \delta_R \rangle_c = 0~,~
\langle \delta_R^2 \rangle_c = \langle \delta_R^2 \rangle \equiv \sigma_R^2~,\cr\cr
&&\langle \delta_R^3 \rangle_c = \langle \delta_R^3 \rangle ~, ~
\langle \delta_R^4 \rangle_c = \langle \delta_R^4 \rangle  - 3 \langle \delta_R^2 \rangle_c^2 ~,
\: {\rm etc.}, 
\end{eqnarray}
with zero mean density field.
Here,
$\sigma_R^2$, $S_3(R)$ and $S_4(R)$ are the variance, the skewness and the
kurtosis, respectively.
Let us consider a non-Gaussian PDF of matter density fluctuations, based on the concept of the Edgeworth expansion.
Here, we consider the expansion of the PDF of the density field $F(\nu)d\nu$ with $\nu \equiv \delta_R / \sigma_R$
in terms of the derivatives of the Gaussian PDF, $F_G(\nu)$, as~\citep{Juszkiewicz:1993hm,LoVerde:2007ri}
\begin{eqnarray}
F(\nu)d\nu = d\nu \left[ c_0 F_G(\nu) + \sum\limits_{m=1} {c_m \over m!} F_G^{(m)}(\nu) \right]~,
\label{eq:PDF}
\end{eqnarray}
with
\begin{eqnarray}
F_G(\nu) &\equiv& (2\pi)^{-1/2} \exp (-\nu^2 / 2)~, \\
F_G^{(m)}(\nu) &\equiv& {d^m \over d\nu^m} F_G(\nu) = (-1)^m H_m(\nu) F_G(\nu)~,
\end{eqnarray}
where $H_m(\nu)$ is the Hermite polynomials;
\begin{eqnarray}
&&H_1(\nu) = \nu~,~H_2(\nu) = \nu^2 -1~,~H_3(\nu) = \nu^3 - 3 \nu~,\nonumber\\
&&H_4(\nu) = \nu^4 - 6 \nu^2 + 3~,~H_5(\nu) = \nu^5 - 10 \nu^3 + 15 \nu~,\cdots~.
\label{eq:Hermite}
\end{eqnarray}
From the above relation between the derivatives of the Gaussian PDF and Hermite polynomials,
we can regard the expression (\ref{eq:PDF}) as a non-Gaussian PDF expanded in terms of
the Hermite polynomials. 
Since the Hermite polynomials satisfy orthogonal relations;
\begin{eqnarray}
\int^{\infty}_{-\infty}
H_m(\nu)H_n(\nu) F_G(\nu) d\nu
=
\left\{
  \begin{array}{cc}
    0   &,~{\rm if}~m \neq n~,    \\
    m!   &,~{\rm if}~m = n~,    \\
  \end{array}
\right.
\end{eqnarray}
we can evaluate the  coefficients as
\begin{eqnarray}
c_m = (-1)^m \int^{\infty}_{-\infty}
H_m(\nu) F(\nu)d\nu~.
\end{eqnarray}
Then, we can obtain the expressions for the
coefficients, $c_m$, in terms of the reduced cumulants
(variance, skewness, kurtosis and so on) as
\begin{eqnarray}
&&c_0=1~,~c_1 = c_2 = 0~,~c_3 = - S_3(R) \sigma_R~,~c_4 = S_4(R) \sigma_R^2~,\cr\cr
&&c_5 = - S_5(R)\sigma_R^3~,c_6= 10 S_3(R)^2\sigma_R^2 + S_6(R)\sigma_R^4~,\cdots~,
\end{eqnarray}
and, as a result,
the non-Gaussian PDF of the density field, $F(\nu)d\nu$, can be obtained as
\begin{eqnarray}
F(\nu)d\nu & = &{d\nu \over \sqrt{2\pi}} \exp \left(-\nu^2 / 2 \right)
\Biggl[
1 + {S_3(R) \sigma_R \over 6}H_3(\nu) + {1 \over 2}\left({S_3(R) \sigma_R \over 6} \right)^2 H_6(\nu)
+{1 \over 6}\left({S_3(R) \sigma_R \over 6} \right)^3 H_9(\nu)
\cr\cr
&&\qquad\quad
+{S_4(R) \sigma_R^2 \over 24}H_4(\nu) + {1 \over 2}\left({S_4(R) \sigma_R^2 \over 24} \right)^2 H_8(\nu)
+
{1 \over 6}\left({S_4(R) \sigma_R^2 \over 24} \right)^3 H_{12}(\nu)
+ \cdots \Biggr]~,
\end{eqnarray}
up to the third order terms in $S_3(R)$ and $S_4(R)$
and neglect the contributions of the higher order cumulants; $S_n(R)$ $(n \ge 5)$.
This derivation of the non-Gaussian PDF is based on the so-called Edgeworth expansion.
Of course, the non-zero non-linearity parameters $f_{\rm NL}$, $\tau_{\rm NL}$ and $g_{\rm NL}$
also generate  non-zero higher order cumulants; $S_n(R)$ $(n \ge 5)$.
However, as far as considering the non-Gaussian curvature fluctuations given by
Eq.~(\ref{eq:multi}) and current observational constraints on
the non-linearity parameters~\citep{Komatsu:2010fb,Fergusson:2010gn},
terms $S_n(R)$ $(n \ge 5)$ are greatly suppressed~\citep{Enqvist:2010bg}.
Hence, the assumption of neglecting the higher order cumulants
seems to be reasonable.

\subsection{Halo mass function with non-Gaussian corrections}
\label{subsec:halomass}

Let us consider the halo mass function with non-Gaussian PDF of the smoothed density field
as given in the previous subsection.
Based on the spirit of the Press-Schechter formula, the halo mass function which gives
the number density of collapsed structures (halos) with the mass between
$M$$(= 4\pi \bar{\rho}R^3/3$ with $\bar{\rho}$ is the background matter density) and
 $M + dM$ at a redshift $z$, $(dn(M,z)/dM) dM$ is given by~\citep{D'Amico:2010ta}
\begin{eqnarray}
{dn \over dM} (M,z) dM &=& -dM{2\bar{\rho} \over M} {d \over dM} \int^{\infty}_{\delta_c / \sigma_R}d\nu F(\nu) \nonumber\\
& = & - dM \sqrt{2 \over \pi} {\bar{\rho} \over M} \exp \left[ - {\nu_c^2 \over 2}\right]
\Biggl\{
{d \ln \sigma_R \over dM}\nu_c
\Biggl[
1 \cr\cr
&&\qquad\qquad
+ {S_3(R)\sigma_R \over 6}H_3(\nu_c)
+ {1 \over 2}\left({S_3(R) \sigma_R \over 6} \right)^2
H_6(\nu_c)
+{1 \over 6}\left({S_3(R) \sigma_R \over 6} \right)^3 H_9(\nu_c)
\cr\cr
&&\qquad\qquad
+{S_4(R) \sigma_R^2 \over 24}H_4(\nu_c) + {1 \over 2}\left({S_4(R) \sigma_R^2 \over 24} \right)^2 H_8(\nu_c)
+
{1 \over 6}\left({S_4(R) \sigma_R^2 \over 24} \right)^3 H_{12}(\nu_c) 
 \Biggr] \cr\cr
&&
\qquad
+
{d\over dM}\left({S_3(R) \sigma_R \over 6} \right) H_2(\nu_c)
+{1 \over 2}{d\over dM}\left({S_3(R) \sigma_R \over 6} \right)^2 H_5(\nu_c) 
+{1 \over 6}{d\over dM}\left({S_3(R) \sigma_R \over 6} \right)^3 H_8(\nu_c) \cr\cr
&&\qquad
 +
{d\over dM}\left({S_4(R) \sigma_R^2 \over 24} \right) H_3(\nu_c)
+{1 \over 2}{d\over dM}\left({S_4(R) \sigma_R^2 \over 24} \right)^2 H_7(\nu_c) 
+{1 \over 6}{d\over dM}\left({S_4(R) \sigma_R^2 \over 24} \right)^3 H_{11}(\nu_c)
\Biggr\} + \cdots~,
\label{eq:nonGaussmass}
\end{eqnarray}
where 
$\nu_c = \delta_c / \sigma_R$ and
$\delta_c$ denotes the threshold for collapse which is originally given by $\delta_c \approx 1.69$.
However, in Ref.~\citep{Grossi:2009an}, the authors have suggested that
using the correction $\delta_c \to \delta_c \sqrt{q}$ with $q=0.75$ puts the analytic predictions in good agreement with the numerical simulations. This is
due to the more realistic case of ellipsoidal collapse. Hence $\delta_c = 1.69 \times \sqrt{q}$ is often referred to 
as the critical density of ellipsoidal collapse.
Here we adopt this corrected density threshold $\delta_c = 1.69 \times \sqrt{0.75}$. 
In the following calculations, we use the above formula of the non-Gaussian mass functions
up to the third order in terms of $S_3$ and $S_4$.

For a Gaussian probability distribution,
the mass function is given by
\begin{eqnarray}
{dn_{\rm G} \over dM} (M,z)dM = 
- \sqrt{2 \over \pi} {\bar{\rho} \over M} \exp \left[ - {\nu_c^2 \over 2}\right]
{d \ln \sigma_R \over dM}
\nu_c dM~,
\label{eq:gaussmass}
\end{eqnarray}
and we define the ratio between the non-Gaussian mass function and the Gaussian one as\footnote{
In Refs.~\citep{Enqvist:2010bg,Matarrese:2000iz,Verde:2000vr},
 the authors introduced the MVJ convention for defining the ratio given by
\begin{eqnarray}
\cal R_{\rm NG}^{\rm MVJ} (M,z)
&=& \exp \left[\nu_c^3 {S_3(R) \sigma_R \over 6} + \nu_c^4 {S_4(R) \sigma_R^2 \over 24} \right]
 \times
\left[ \delta_3 + {\nu_c \over \delta_3}
\left( - {S_3(R) \sigma_R \over 6} \right) 
+ \left({ d \ln \sigma_R \over dM} \right)^{-1}
{d \over dM}\left( {S_3(R) \sigma_R \over 6} \right)\right]
\cr\cr
&& \qquad\qquad \times
\left[ \delta_4 + {\nu_c^2 \over \delta_4}
\left( - {S_4(R) \sigma_R^2 \over 12} \right) 
+ \left({ d \ln \sigma_R \over dM} \right)^{-1}
{d \over dM}\left( {S_4(R) \sigma_R^2 \over 24} \right)\right]~, \cr\cr
\delta_3
& \equiv &
\left( 1 - \nu_c {S_3(R) \sigma_R \over 3} \right)^{1/2}~,~
\delta_4
 \equiv
\left( 1 - \nu_c^2 {S_4(R) \sigma_R^2 \over 12} \right)^{1/2}~,
\end{eqnarray}
which is not based on the Edgeworth expansion.
In our calculation, we have also checked the consistency between the above MVJ expression
and Eq. (\ref{eq:ratio}). 
This issue is discussed in Appendix~\ref{app:mvj}.
}

\begin{eqnarray}
{\cal R}_{\rm NG}(M,z) \equiv {dn(M,z)/dM \over dn_{\rm G}(M,z)/dM}~.
\label{eq:ratio}
\end{eqnarray}
Let us focus on the redshift dependence of the above expression.
From the definition of the reduced cumulants~(\ref{eq:cum}) and the 
fact that the
redshift dependence of the density field is given by $\delta_R \propto D(z)$,
we can easily find that $\sigma^{p-2}S_p(R)$ has no redshift-dependence.
Hence,
any remaining redshift dependence comes only from the term ${\delta_c \over \sigma_R}$.
Here, following the literature,
the redshift-dependence can be carried by $\delta_c$ as $\delta_c \to \delta_c(z) \propto D(z)^{-1}$
and then the variance $\sigma_R$ has no redshift-dependence. 
In the following discussion, we change the subscript $R$ to $M$ because $R$ and $M$ have a one-to-one correspondence
through the equation $M = 4\pi R^3 \bar{\rho}/3$.

\subsubsection{Variance, skewness and kurtosis}
\label{subsubsec:vsk}

\begin{figure}
  \begin{center}
    \includegraphics{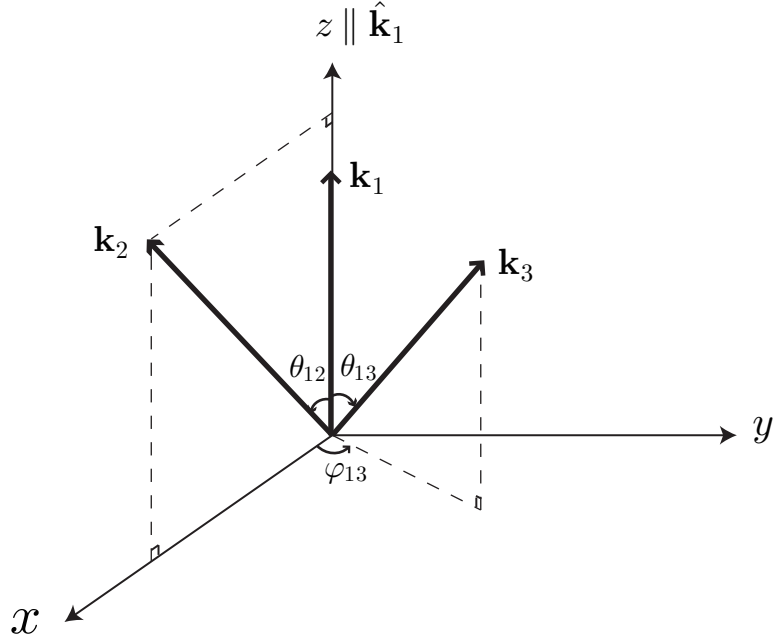}
  \end{center}
  \caption{the three vectors, ${\bf k}_1$, ${\bf k}_2$ and ${\bf k}_3$ in the trispectrum.}
  \label{fig:4_point.eps}
\end{figure}

Let us consider the concrete expressions of the variance, skewness and kurtosis
of the primordial curvature perturbations whose power-, bi- and tri-spectra
are given by Eqs. (\ref{eq:powerspectrum}), (\ref{eq:bispectrum})
and (\ref{eq:trispectrum}), respectively.
The variance is given by
\begin{eqnarray}
\sigma_R^2 = \int {dk \over k} W_R^2(k){\cal M}(k)^2{\cal P}_\phi(k) ~, 
\label{eq:fullvariance}
\end{eqnarray}
the skewness is~\citep{Chongchitnan:2010xz}
\begin{eqnarray}
 S_3(R) &\equiv& {6 \over 5}{f_{\rm NL} \over \sigma_R^4} \tilde{S}_3(R)~, \nonumber\\
 \tilde{S}_3(R)&=& 
 \int { d k_1 \over k_1}
 W_R(k_1){\cal M}(k_1) {\cal P}_\phi(k_1) 
\int { d k_2 \over k_2} 
W_R(k_2){\cal M}(k_2){\cal P}_\phi(k_2)~\nonumber\\
&& 
\qquad
\times
\int {d \mu_{12} \over 2} W_R(k_{12}){\cal M}(k_{12})
\biggl[ 1 + {P_\phi(k_{12}) \over P_\phi(k_1)} + {P_\phi(k_{12}) \over P_\phi(k_2)}
\biggr]~,
 \label{eq:fullskewness}
\end{eqnarray}
where $k_{12} = \sqrt{k_1^2 + k_2^2 + 2k_1k_2 \mu_{12}}$ and $\mu_{12} = \cos \theta_{12}$,
and the kurtosis which is proportional to the non-linearity parameter $\tau_{\rm NL}$ is
given by
\begin{eqnarray}
 S_4^{\tau}(R) & \equiv & {\tau_{\rm NL} \over \sigma_R^6} \tilde{S}_4^{\tau}(R)~, \nonumber\\
 \tilde{S}_4^{\tau}(R) &=&
  \int {dk_1 \over k_1}W_R(k_1){\cal M}(k_1){\cal P}_\phi (k_1) 
 \int {dk_2 \over k_2}W_R(k_2){\cal M}(k_2){\cal P}_\phi (k_2)
\int { dk_3 \over k_3}W_R(k_{3}){\cal M}(k_3) {\cal P}_\phi (k_3)
\cr\cr
&& 
\times   \int^{1}_{-1} {d\mu_{12} \over 2}
\int^{1}_{-1} {d\mu_{13} \over 2}
 \int^{2\pi}_0 {d\varphi_{13} \over 2\pi} W_R(k_{123}){\cal M}(k_{123})
\cr\cr
&&
\times
 \Biggl\{P_{\phi}(k_{12}) \left[ {1 \over P_\phi(k_1)} + {1 \over P_\phi(k_2)} \right]
 \left[ 1 +  {P_\phi(k_{123}) \over P_\phi(k_3)} \right] 
\cr\cr
&&
+ P_{\phi}(k_{23}) \left[ {1 \over P_\phi(k_2)} + {1 \over P_\phi(k_3)} \right]
 \left[ 1 +  {P_\phi(k_{123}) \over P_\phi(k_1)} \right]
+ P_{\phi}(k_{31}) \left[ {1 \over P_\phi(k_3)} + {1 \over P_\phi(k_1)} \right]
 \left[ 1 +  {P_\phi(k_{123}) \over P_\phi(k_2)} \right] \Biggr\}~.
 \label{eq:fullkurtosis}
\end{eqnarray}
Here, we have fixed the three vectors, ${\bf k}_1$, ${\bf k}_2$ and ${\bf k}_3$
that appear in the expression of the trispectrum, as shown in Fig.~\ref{fig:4_point.eps}.
Hence, using the angular variables, $\theta_{12}$, $\theta_{13}$ and $\varphi_{13}$, 
we have
\begin{eqnarray}
&&
k_{12} = \sqrt{k_1^2 + k_2^2 + 2 k_1k_2\mu_{12}}~,\cr\cr
&&
k_{23} = \sqrt{k_2^2 + k_3^2 + 2k_2 k_3 \left( \sqrt{(1 - \mu_{12}^2)(1 - \mu_{13}^2)} \cos \varphi_{13} + \mu_{12}\mu_{13}\right)}~, \cr\cr
&&
k_{13} = \sqrt{k_1^2 + k_3^2 + 2 k_1k_3\mu_{13}}~,
\end{eqnarray}
and 
\begin{eqnarray}
k_{123} = \sqrt{k_1^2 + k_2^2 + k_3^2 + 2k_1k_2\mu_{12}+2k_1k_3\mu_{13}
+2k_2k_3\left( \sqrt{(1 - \mu_{12}^2)(1 - \mu_{13}^2)} \cos \varphi_{13} + \mu_{12}\mu_{13}\right)}~,
\end{eqnarray}
where $\mu_{ij} \equiv \cos \theta_{ij}$.
In order to calculate the skewness more easily,
let us
consider the
squeezed limit in  momentum space, e.g., $k_1 \ll k_2 \simeq k_3$.
In this limit, the equation for the skewness (\ref{eq:fullskewness}) can be reduced to
\begin{eqnarray}
\tilde{S}_3 \biggr|_{k_1 \ll k_2 \simeq k_3} \simeq 2 \sigma_R^2
 \int {dk_1 \over k_1}W_R(k_1){\cal M}(k_1){\cal P}_\phi (k_1)~,
\end{eqnarray}
and by considering other limiting cases, i.e., $k_2 \to 0$ and $k_3 \to 0$,
we obtain 
\begin{eqnarray}
\tilde{S}_3  \simeq 6 \sigma_R^2
 \int {dk \over k}W_R(k){\cal M}(k){\cal P}_\phi (k)~.
 \label{eq:appskew}
\end{eqnarray}
Based on the above approximate expression, we  find a simple formula;
\begin{eqnarray}
\sigma_R S_3(R) = 4.3 \times 10^{-4} f_{\rm NL} \times \sigma_R^{0.13}~~( 10^{12} h^{-1} M_{\odot} < M < 2 \times 10^{15} h^{-1} M_{\odot})~.
\label{eq:anaskew}
\end{eqnarray}
This result seems to be close to those given in Refs.~\citep{DeSimone:2010mu,Enqvist:2010bg}\footnote{
As mentioned in Ref.~\citep{Enqvist:2010bg}, this result is different from that in the published version of Ref.~\citep{Chongchitnan:2010xz}.
However, the authors in Ref.~\citep{Chongchitnan:2010xz} have corrected the result in the arXiv version and
their new derivation is now  close to our result~(\ref{eq:anaskew}). 
}.
Hence, we adopt the above expression in the following discussion.
In a similar way,
from the expression of the kurtosis (\ref{eq:fullkurtosis}),
we can easily find that
the kurtosis induced from the non-linearity parameter $\tau_{\rm NL}$
becomes largest in the limit of $k_i \to 0 (i=1,2,3,4)$ or $k_{ij} \to 0 (i \neq j=1,2,3,4)$ (local type).
Then, we have an approximate expression 
\begin{eqnarray}
\tilde{S}_4^{\tau}(R) \simeq 8 \int
{dk \over k} W_R(k){\cal M}(k) {\cal P}_\phi (k) \times \tilde{S}_3(R) + 12 A_\phi \sigma_R^4~.\nonumber\\
\label{eq:approxkurt}
\end{eqnarray}
On the other hand, in the squeezed limit $k_i \to 0 (i=1,2,3,4)$, the kurtosis which is
proportional to the non-linearity parameter $g_{\rm NL}$
can be also reduced to~\citep{Chongchitnan:2010xz,Enqvist:2010bg}
\begin{eqnarray}
S_4^{g} &\equiv& {54 \over 25}{ g_{\rm NL} \over \sigma_R^6}  \tilde{S}_4^{g}~,\cr\cr
\tilde{S}_4^{g} &\simeq& 2 \int
{dk \over k} W_R(k){\cal M}(k) {\cal P}_\phi (k) \times \tilde{S}_3(R)~.
\end{eqnarray}
From these approximate expressions, we respectively obtain simple formulae for the kurtosis in the form
\begin{eqnarray}
\sigma_R^2 S_4^{\tau}(R) &=& 1.9 \times 10^{-7} \tau_{\rm NL} \times \sigma_R^{0.25}~~( 10^{12} h^{-1} M_{\odot} < M < 2 \times 10^{15} h^{-1} M_{\odot})~, \cr\cr
\sigma_R^2 S_4^g(R) &=& 9.4 \times 10^{-8} g_{\rm NL} \times \sigma_R^{0.27}~~( 10^{12} h^{-1} M_{\odot} < M < 2 \times 10^{15} h^{-1} M_{\odot})~.
\label{eq:anakurt}
\end{eqnarray}
The result for $S_4^{g}(R)$ also is close to that obtained in Ref.~\citep{Enqvist:2010bg}\footnote{
In addition to the expression  derived for the skewness, the result for kurtosis  in Ref.~\citep{Chongchitnan:2010xz} has been corrected in the arXiv version 
and this is also close to our result~(\ref{eq:anakurt}).}.
Hence in the following discussion ,
we also adopt the above expressions for the kurtosis as well as that for the skewness.


\subsubsection{Difference between the Gaussian and the non-Gaussian mass functions}
\label{subsub:dif}

Based on the above calculations for the variance, $\sigma_R^2$, the skewness, $S_3$, and also the kurtosis, $S_4$,
the mass function can now be calculated.
In the following discussion,
we take values of the non-linearity parameters
as $f_{\rm NL} = 100$, $\tau_{\rm NL} = 10^6$ and $g_{\rm NL} = 0$.
This value of $\tau_{\rm NL}$ may be inconsistent with the observational constraint obtained by Ref.~\citep{Smidt:2010ra}
as $-0.6 < \tau_{\rm NL} / 10^4 < 3.3$ at $95\%$ confidence level.
However, there might be a caveat since in Ref.~\citep{Fergusson:2010gn},
the authors have claimed
that the approach in Smidt et al.
does not directly subtract the effect of anisotropic noise and other
systematic effects which are important in obtaining an accurate and optimized
result.
Nonetheless,
in order to emphasize the differences between the Gaussian mass functions and the non-Gaussian mass functions
with the non-zero $f_{\rm NL}$ and the non-zero $\tau_{\rm NL}$ cases,
we take the above values.

\begin{figure}
 \begin{center}
    \includegraphics{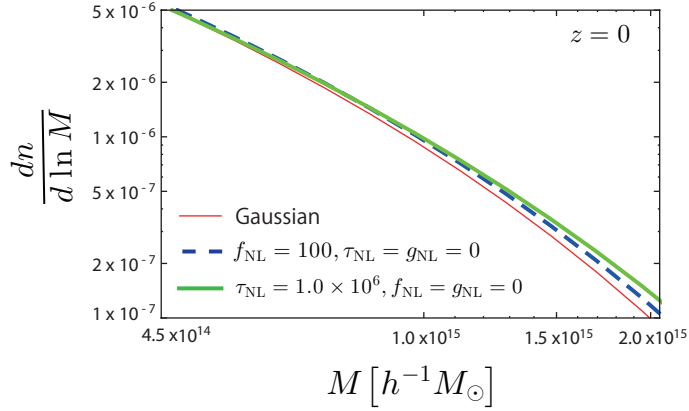}
  \end{center}
  \caption{The mass function with the mass range between $4.5 \times 10^{14} h^{-1} M_{\odot}$ and $2.0 \times 10^{15} h^{-1} M_\odot$ at the redshift $z=0$.
  The red thin line shows the mass function with the Gaussian density fluctuations given by Eq.~(\ref{eq:gaussmass}). The blue dashed and green thick lines show
  the non-Gaussian mass function given by Eq.~(\ref{eq:nonGaussmass}) 
  in the case with $f_{\rm NL}=100$ and $\tau_{\rm NL} = g_{\rm NL} = 0$ and the case with $\tau_{\rm NL} = 10^6$ and $f_{\rm NL} = g_{\rm NL} = 0$, respectively.}
  \label{fig:mass_func_0.eps}
\end{figure}

\begin{figure}
 \begin{center}
    \includegraphics{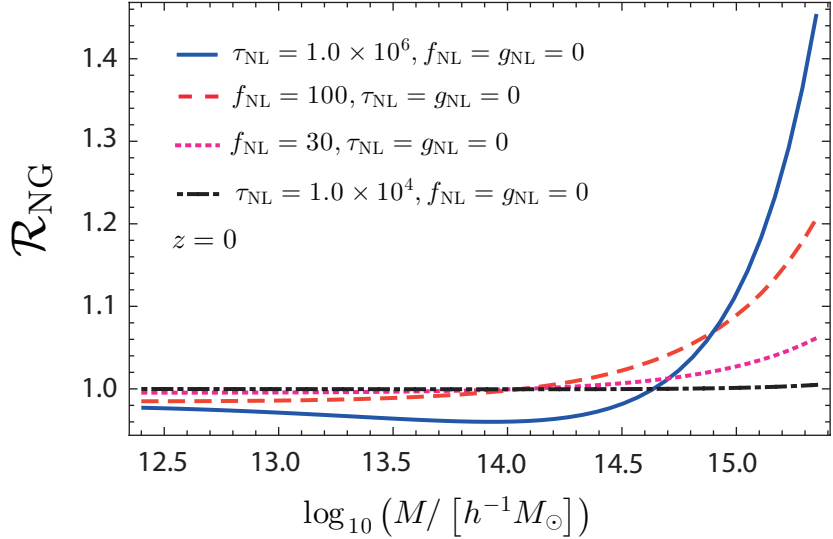}
  \end{center}
  \caption{The ratio between the Gaussian and the non-Gaussian mass functions.
   The red dashed line shows ${\cal R}_{\rm NG}(M,0)$ with $f_{\rm NL} = 100$ and $\tau_{\rm NL} = g_{\rm NL} = 0$
   and the blue solid line shows that with $f_{\rm NL} = g_{\rm NL} = 0$ and $\tau_{\rm NL} = 10^6$.
   The magenta dotted line is for the case with $\tau_{\rm NL} = g_{\rm NL} = 0$ and $f_{\rm NL} = 30$
   which is corresponding to the mean value of the current WMAP data~\citep{Komatsu:2010fb}.
   The black dashed-dotted line is for the case with $f_{\rm NL} = g_{\rm NL} = 0$ and $\tau_{\rm NL} = 10^4$
   which is consistent with the maximum allowed value obtained by Ref.~\citep{Smidt:2010ra}.}
  \label{fig:mass_func_ratio.eps}
\end{figure}

In Fig.~\ref{fig:mass_func_0.eps},
we show that the mass function in  the mass range between $5.0 \times 10^{14} h^{-1} M_{\odot}$ and $2.0 \times 10^{15} h^{-1} M_\odot$ at the redshift $z=0$.
  The red thin line shows the mass function with the Gaussian density fluctuations given by Eq.~(\ref{eq:gaussmass}). The blue dashed and green thick lines show
  the non-Gaussian mass function given by Eq.~(\ref{eq:nonGaussmass}) 
  in the cases with $f_{\rm NL}=100$ and $\tau_{\rm NL} = g_{\rm NL} = 0$ and  $\tau_{\rm NL} = 10^6$ and $f_{\rm NL} = g_{\rm NL} = 0$, respectively.
From this figure, it is rather difficult to see the differences between the Gaussian and the non-Gaussian mass functions.
In Fig.~\ref{fig:mass_func_ratio.eps} we show the ratios between the Gaussian and the non-Gaussian mass functions
defined by Eq.~(\ref{eq:ratio}).
The red dashed line shows ${\cal R}_{\rm NG}(M,0)$ with $f_{\rm NL} = 100$ and $\tau_{\rm NL} = g_{\rm NL} = 0$
and the blue solid line shows that with $f_{\rm NL} = g_{\rm NL} = 0$ and $\tau_{\rm NL} = 10^6$.
The magenta dotted line is for the case with $\tau_{\rm NL} = g_{\rm NL} = 0$ and $f_{\rm NL} = 30$
    which is corresponding to the mean value of the current WMAP data~\citep{Komatsu:2010fb}.
   The black dashed-dotted line is for the case with $f_{\rm NL} = g_{\rm NL} = 0$ and $\tau_{\rm NL} = 10^4$ which
   is consistent with the maximum allowed value obtained by Ref.~\citep{Smidt:2010ra}.%
\begin{figure}
  \begin{center}
    \includegraphics{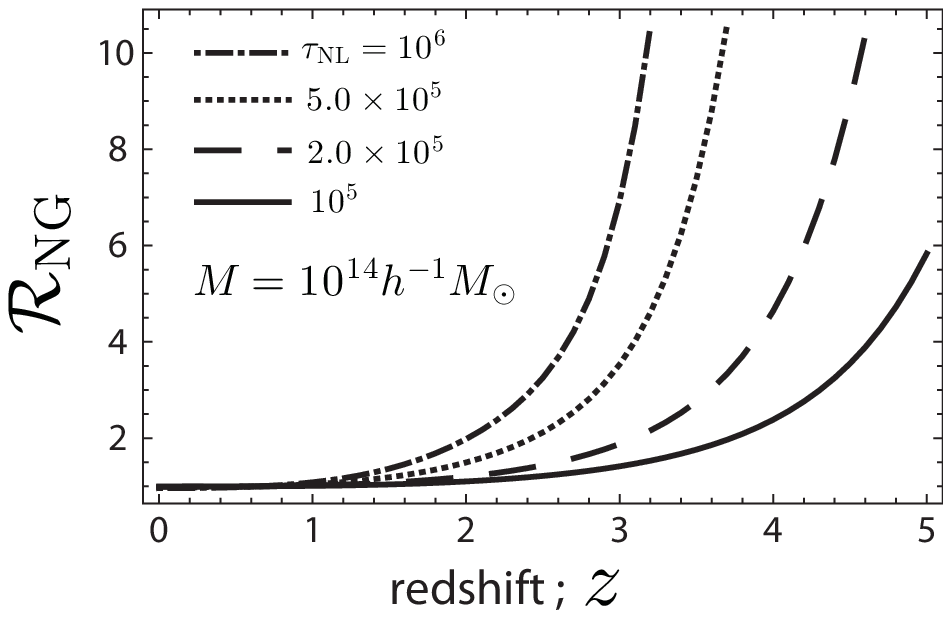}
  \end{center}
  \caption{The ratio between the non-Gaussian mass function and the Gaussian mass function
  v.s. the redshift ($0 < z < 5$) with changing the value of $\tau_{\rm NL}$.
  Here we have fixed the mass of halo as $M = 10^{14} h^{-1} M_{\odot}$.
  The solid line is for the case with $\tau_{\rm NL} = 10^5$,
  dashed line for $2.0 \times 10^5$, dotted line for $5.0 \times 10^5$ and dot-dashed line for $10^6$.}
  \label{fig:mass_func_z.eps}
\end{figure}
From this figure, we  infer that
for both types of primordial non-Gaussianity, i.e.,  positive skewness and kurtosis,
the mass functions can be systematically enhanced for more massive objects, as  compared with the Gaussian case.
The enhancement of the mass functions depends on the values of $\tau_{\rm NL}$ and $g_{\rm NL}$.
We find that for the cases with $f_{\rm NL}=30$ and $\tau_{\rm NL} = 10^4$
${\cal R}_{\rm NG}$ are respectively $1.06$ and $1.01$ for $M = 2 \times 10^{15}  h^{-1} M_{\odot} $.
Hence, in both cases the effects of the primordial non-Gaussianity on the mass functions
seem to be too small to detect.
We  also find that the enhancement of
the non-Gaussian mass function with the  non-zero kurtosis type of  primordial non-Gaussianity,
i.e., non-zero $\tau_{\rm NL}$,
depends more strongly  on the mass of the collapsed objects than
the case with the non-zero skewness type of primordial non-Gaussianity.
This is because 
in the expression for the non-Gaussian mass function (\ref{eq:nonGaussmass}), the
$\delta_c/\sigma_R$-dependence of the term related with the kurtosis $S_4$
is stronger than that of the term related with the skewness $S_3$,
namely, $S_4$-term $\propto (\delta_c/\sigma_R)^{5}$ and $S_3$-term
$\propto (\delta_c/ \sigma_R)^4$. 
As the collapsed objects become more massive,
the variance $\sigma_R$ becomes smaller and hence
$\delta_c / \sigma_R$ becomes larger.
Thus, if we would detect the enhancement of the mass function for
massive collapsed objects and find its scale-dependence,
then we might distinguish  the kurtosis type of  primordial non-Gaussianity 
from the skewness type.
In Fig.~\ref{fig:mass_func_z.eps},
we show the redshift dependence of the ratio between the non-Gaussian mass function
and the Gaussian mass function as we  change the value of $\tau_{\rm NL}$.
Here we have fixed the mass of the halo as $M = 10^{14} h^{-1} M_{\odot}$.
The solid line is for the case with $\tau_{\rm NL} = 10^5$,
the dashed line for $2.0 \times 10^5$, dotted line for $5.0 \times 10^5$ and the dashed-dotted line for $10^6$.
From this figure, we find that at higher redshift
the enhancement of the mass function for  massive collapsed objects increases.
This is because the critical density $\delta_c(z) = \delta_c / D(z)$ becomes
much larger at larger redshifts due to the smaller linear growth function $D(z)$.
Hence, in order to observationally test the kurtosis type of primordial non-Gaussianity
it will be useful to observe high-redshift rare objects.

\section{Applications}
\label{sec:nGappl}

In this section,
we consider applications of the mass function with both
skewness and kurtosis types of primordial non-Gaussianity.
Here, we also take values of the non-linearity parameters
to be  $f_{\rm NL} = 100$, $\tau_{\rm NL} = 10^6$ and $g_{\rm NL} = 0$.


\subsection{Early Star Formation}
\label{sub:early}

Let us first investigate 
the effect of primordial non-Gaussianity
on the epoch of reionization.
As is well-known,
in order to understand the mechanism of  reionization,
it is important to estimate the number of photons from Population III stars.
Following Refs.~\citep{Somerville:2003sk,Somerville:2003sh,Sugiyama:2003tc},
the global star-formation-rate density denoted by $\dot{\rho}_*$
can be written as 
\begin{eqnarray}
\dot{\rho}_* = e_* \rho_b {d \over dt}
F_h (M_{\rm vir} > M > M_{\rm crit},t)~. 
\end{eqnarray}
Here, $\rho_b$ is the background baryon number density
and $e_*$ denotes the star-formation efficiency
usually
taken to be 0.002
for $200 M_{\odot }$ Pop III stars
and 0.001 for $ 100 M_{\odot}$.
$F_h (M_{\rm vir} > M > M_{\rm crit},t)$
represents the fraction of the total mass
in collapsed objects (halos) with
masses greater than the minimum collapse mass scale $M_{\rm crit} = 10^6
h^{-1} M_{\odot}$~\citep{Yoshida:2003rw,Fuller:2000vk}
and lower than the virial mass $M_{\rm vir} = M(T_{\rm vir} = 10^4 {\rm K})$.

The relation between the mass and the virial temperature  
is given by~\citep{Barkana:2000fd,Yoshida:2003rw}
\begin{eqnarray}
T_{\rm vir}
= 4.7 \times 10^3 \left( {\mu \over 0.6} \right)
\left( {M \over 10^8 h^{-1} M_{\odot}} \right)^{2/3}
\left( {\Omega_{m0} \over 0.24}{\Delta_c(z) \over 18 \pi^2}\right)^{1/3}
\left( {1+z \over 10} \right)~{\rm K}
\end{eqnarray}
where $\mu$ is the mean molecular weight,
and
$\Delta_c(z)$ is the final overdensity relative to the critical
density, which is given by a fitting formula~\citep{Bryan:1997dn}
\begin{eqnarray}
\Delta_c = 18 \pi^2 + 82 \left( \Omega_m(z) - 1\right)
-39 \left( \Omega_m(z) - 1\right)^2~,
\end{eqnarray}
where
$\Omega_{m}(z)$ is
the density parameter of matter at  redshift $z$; 
\begin{eqnarray}
\Omega_m(z) = {\Omega_{m0}(1+z)^3 \over \Omega_{m0}(1+z)^3 + \Omega_{\Lambda}}~.
\end{eqnarray}
Assuming that the photon number production-rate per $ M_{\odot}$ from
Pop. III stars is $N_{\gamma} = 1.6
\times 10^{48} s^{-1} M_{\odot}^{-1}$ and that the life time of Pop. III star
is $\tau_{\rm III} = 3.0 \times 10^6 {\rm yr}$, we can obtain the total production rate of
ionizing photons at  time $t$ as
\begin{eqnarray}
{d n_\gamma \over dt}(t) = e_* \rho_b
N_{\gamma} \left(F_h(t) - F_h(t - \tau_{\rm III})\right)~,
\end{eqnarray}
hence the cumulative number of photons per H atom is
\begin{eqnarray}
{n_\gamma \over n_H}(z) \simeq
\mu m_p e_* N_\gamma F_h(M_{\rm vir} > M > M_{\rm crit},z)\tau_{\rm III}~,
\label{eq:cumnum}
\end{eqnarray}
with the proton mass $m_p$ and the hydrogen number density
$n_H$.
In the above expression,
$F_h (M_{\rm vir} > M > M_{\rm crit},z)$ is given by Press-Schechter theory as
\begin{eqnarray}
F_h (M_{\rm vir} > M > M_{\rm crit}, z)
= {1 \over \bar{\rho}} \int^{M_{\rm vir}}_{M_{\rm crit}}
M {d n \over dM}(M,z)dM~.
\end{eqnarray}
%

Substituting our expression~(\ref{eq:nonGaussmass}) 
for the non-Gaussian mass function into the above equation,
we can estimate the effect of primordial non-Gaussianity
on the number of photons emitted from Population III stars,
which is one of the most  important quantities during the epoch
of reionization. 
\begin{figure}
  \begin{center}
    \includegraphics[width=85mm]{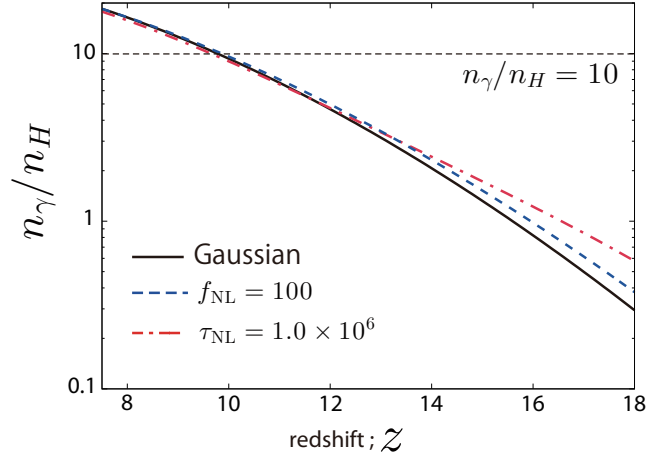}
  \end{center}
  \caption{
  Cumulative photon number per $H$ atom as a function of the redshift for $8 < z < 18$.
  The black solid line shows $n_\gamma / n_H (z)$ for the case with the Gaussian fluctuations.
  The blue dashed line is for the case with the non-Gaussian fluctuations; $f_{\rm NL} = 100$ and $\tau_{\rm NL} = g_{\rm NL} = 0$.
  The red dot-dashed line is for the case with $\tau_{\rm NL} = 1.0 \times 10^6$ and $f_{\rm NL} = g_{\rm NL} = 0$.
  The thin black dashed line corresponds to $n_\gamma / n_H$ as a guide of the complete reionization. 
  }%
  \label{fig:photon_H_ratio.eps}
\end{figure}
In Fig.~\ref{fig:photon_H_ratio.eps}, we show the cumulative photon number per $H$ atom given by Eq.~(\ref
{eq:cumnum}) as a function of the redshift for $ 8 < z < 18$.
The black solid line shows $n_\gamma / n_H (z)$ for the case with the Gaussian fluctuations,
the blue dashed line is for the case with the non-Gaussian fluctuations; $f_{\rm NL} = 100$ and $\tau_{\rm NL} = g_{\rm NL} = 0$ and
the red dot-dashed line for the case with $\tau_{\rm NL} = 1.0 \times 10^6$ and $f_{\rm NL} = g_{\rm NL} = 0$.
The thin black dashed line corresponds to $n_\gamma / n_H = 10$
as a guide of the complete reionization on average.
From this figure,
we find that primordial non-Gaussianity 
seems not to affect the reionization history of the Universe
on average which is characterized by the value of
$n_\gamma / n_H = 10$.
However, at  higher redshift
the effect of primordial non-Gaussianity
seems to be significant
on the cumulative photon number density.
We 
evaluate   this effect
\begin{figure}
  \begin{center}
    \includegraphics{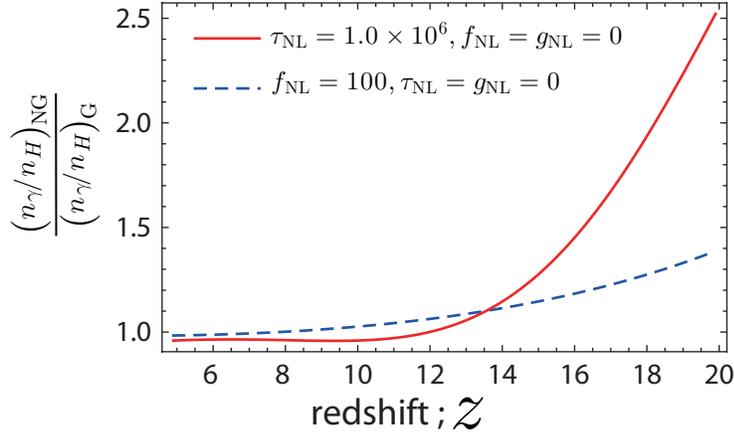}
  \end{center}
  \caption{We plot the ratio between
$n_\gamma (z)/ n_H$
in the pure Gaussian primordial fluctuation case
and that in the non-Gaussian case
for $5 < z < 20$.
The blue dashed line is for the case with
$f_{\rm NL} = 100$ and $\tau_{\rm NL} = g_{\rm NL} = 0$ and
the red solid line for the case with $ \tau_{\rm NL}
= 1.0 \times 10^6$ and $f_{\rm NL} = g_{\rm NL} = 0$.
}
  \label{fig:photon_number.eps}
\end{figure}
in Fig.~\ref{fig:photon_number.eps}, where
we plot the ratio between
$n_\gamma / n_H (z)$
in the pure Gaussian primordial fluctuation case
and that in the non-Gaussian case.
The blue dashed line is for the case with
$f_{\rm NL} = 100$ and $\tau_{\rm NL} = g_{\rm NL} = 0$ and
the red solid line for the case with $ \tau_{\rm NL}
= 1.0 \times 10^6$ and $f_{\rm NL} = g_{\rm NL} = 0$.
From this figure, we find that
compared to the Gaussian case
the cumulative number of photons in the
non-Gaussian case is larger at higher redshifts
both in the non-zero $S_3$ and the non-zero $S_4$ cases.
Moreover, as we have mentioned in the previous section
the kurtosis type of primordial non-Gaussianity
affects
the enhancement of the photon number density
more significantly at  high redshift.
That is,
there seems to be the possibility of 
dramatically changing the
history of the early stage of reionization
due to the kurtosis type of primordial non-Gaussianity
even for values in the range of the current
limits obtained from CMB observations.
Of course, the above rough estimate
is not precise enough to enable us to estimate the exact cumulative number of the ionizing photons.
However, we consider here that, in view of the completely ad hoc nature of the amount of non-Gaussianity due to the absence of a compelling inflationary model,  it suffices for us 
to focus on the deviation of the photon number
based on the non-Gaussian mass function from that based on the Gaussian mass function.

\subsection{High-Redshift Massive Clusters}
\label{sub:mass}

Recently, 
the authors in Ref.~\citep{Jee:2009nr,Rosati:2009cm} have presented a weak lensing analysis of the galaxy cluster XMMU J2235.3-2557 which has
a high redshift $z \approx 1.4$ and whose mass is
$M_{324} = (6.4 \pm 1.2) \times 10^{14} M_{\odot}$\footnote{
The halo is defined as a spherical overdense region
whose density is 324 times the mean matter density of the Universe.
}.
In $\Lambda$CDM model
the formation of such a massive cluster at this redshift would be a rare event (at least $3\sigma$).

In Ref.~\citep{Cayon:2010mq},
the authors have considered
the effects of primordial non-Gaussianity parametrized by the non-linearity parameter
$f_{\rm NL}$ which they found to be 
$f_{\rm NL} = 449 \pm 286$ at wave number of about $0.4 {\rm Mpc}^{-1}$
in order to explain the existence of such a massive cluster
at high redshift.
Considering scale-invariant $f_{\rm NL}$,
this result contradicts
the current CMB observational constraint  $f_{\rm NL} < 100$.
Therefore,
the authors remarked that one would need to invoke  scale-dependent $f_{\rm NL}$.
In Ref.~\citep{Enqvist:2010bg},
the authors have considered  non-zero $g_{\rm NL}$ case
and found that $g_{\rm NL} = O(10^6)$
could explain the existence of high redshift massive clusters.

Here, instead of considering the scale-dependence of $f_{\rm NL}$ or $g_{\rm NL}$,
let us consider the effect of the kurtosis induced from the $\tau_{\rm NL}$-type
primordial non-Gaussianity on the formation of massive clusters.
Of course, for a more detailed analysis
we need to calculate the
probability of the massive clusters under the procedure done in Ref.~\citep{Cayon:2010mq}.
However, 
in order to give a naive estimation of the value of $\tau_{\rm NL}$ which can explain the
existence of the massive cluster XMMU J2235.3-2557,
we investigate the value of $\tau_{\rm NL}$
which gives the same value as does  the non-Gaussian mass function, namely,
${\cal R}_{\rm NG}$ defined as Eq.~(\ref{eq:ratio}),
including the effect of kurtosis $S_4$
 on the corresponding scale at the corresponding
redshift by including the effect of  $f_{\rm NL}$, i.e., skewness.
Here, we
adopt $M_{\rm XMMU} = 6.4 \times 10^{14} M_{\odot}$
and $z_{\rm XMMU} = 1.4$ as
the mass and the redshift
of the massive cluster XMMU J2235.3-255, respectively.
For the value of $f_{\rm NL}$, 
the best fit value derived in Ref.~\citep{Cayon:2010mq}
is adopted.
For these parameters, 
we also find that
this value can be realized in the case with
$f_{\rm NL} = g_{\rm NL} = 0$ and $\tau_{\rm NL} = 1.7 \times 10^6$.
As we have mentioned in Sec. \ref{subsub:dif},
this value may be ruled out by the result obtained by
Ref.~\citep{Smidt:2010ra}.
Hence, if we believe this constraint,
we need to consider the possibility such as scale-dependent $\tau_{\rm NL}$.

\subsection{Abundance of voids}
\label{sub:void}

As another example,
we study
the void abundance with primordial non-Gaussian corrections.
In Ref.~\citep{Kamionkowski:2008sr},
the authors showed that
the void distribution function can be derived in the same way as the halo
mass function using  Press-Schechter theory.
This is done by replacing
the critical "overdensity" parameter, $\delta_c$,
with the negative 
"underdensity" parameter, $\delta_v$.
The precise value of $\delta_v$ depends on the definition of a void.
For example, if the voids are regions having a density half of $\bar{\rho}$~,
then we can estimate the critical value of underdensity as $\delta_v \simeq -0.7$~\citep{Kamionkowski:2008sr}.
There are also several numerical studies about the value of $\delta_v$ which suggest $\delta_v \approx -0.8$~\citep{Shandarin:2005ea,Park:2006wu,Colberg:2008qg}.

\begin{figure}
  \begin{center}
    \includegraphics{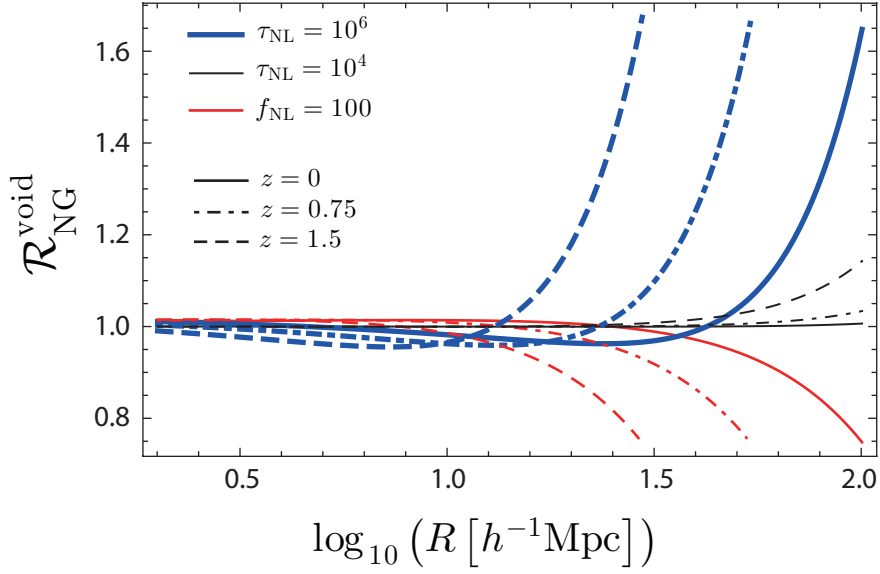}
  \end{center}
  \caption{The ratio between the void abundance distribution function with non-zero
  primordial non-Gaussian correction and the Gaussian distribution function at redshifts $z=0.0$, $0.75$, $1.5$
  given by Eq.~(\ref{eq:voidratio}).
  The red lines are for the case with $f_{\rm NL} = 100$ and $\tau_{\rm NL} = g_{\rm NL} = 0$,
  the blue thick lines for the case with $\tau_{\rm NL} = 10^6$ and $f_{\rm NL} = g_{\rm NL} = 0$
  and the black thin lines for the case with $\tau_{\rm NL} = 10^4$ and $f_{\rm NL} = g_{\rm NL} = 0$.
  The solid lines are for the case with $z=0.0$, the dot-dashed lines for the case with $z=0.75$ and the dashed lines for the case with $z=1.5$.
  Here we took $\delta_v = -0.7$.}
  \label{fig:void_ratio.eps}
\end{figure}

In any case, based on Press-Schechter theory, the abundance of voids which have radius between $R$ and $R + d R$ is
given by~\citep{Kamionkowski:2008sr}
\begin{eqnarray}
{dn^{\rm void}(R) \over dR} dR = - dR \times  {6 \over 4 \pi R^3} 
{d \over dR}\int_{-\infty}^{\delta_v / \sigma_R} F(\nu)d\nu ~.
\end{eqnarray}
For pure Gaussian PDF, we have
\begin{eqnarray}
{dn_{\rm G}^{\rm void}(R) \over dR}
= \sqrt{2 \over \pi} {3 \over 4 \pi R^4} \exp \left[ -{\delta_v^2 \over 2 \sigma_R^2}\right] 
 {\delta_v \over \sigma_R} {d \ln \sigma_R \over d \ln R}~.
\end{eqnarray}
Up to the third order in terms of $S_3$ and $S_4$, the void abundance with primordial non-Gaussian corrections
is also given by
\begin{eqnarray}
{dn^{\rm void} (R) \over dR}
&=& \sqrt{2 \over \pi} {3 \over 4 \pi R^4} \exp \left[ -{\delta_v^2 \over 2 \sigma_R^2}\right] 
\Biggl\{ {d \ln \sigma_R \over d \ln R} {\delta_v \over \sigma_R}
\Biggl[
1 \cr\cr
&&\qquad
+ {S_3(R)\sigma_R \over 6}H_3(\delta_v / \sigma_R)
+ {1 \over 2}\left({S_3(R) \sigma_R \over 6} \right)^2
H_6(\delta_v / \sigma_R)
+{1 \over 6}\left({S_3(R) \sigma_R \over 6} \right)^3 H_9(\delta_v / \sigma_R)
\cr\cr
&&\qquad
+{S_4(R) \sigma_R^2 \over 24}H_4(\delta_v / \sigma_R) + {1 \over 2}\left({S_4(R) \sigma_R^2 \over 24} \right)^2 H_8(\delta_v / \sigma_R)
+
{1 \over 6}\left({S_4(R) \sigma_R^2 \over 24} \right)^3 H_{12}(\delta_v / \sigma_R) 
 \Biggr] \cr\cr
&&
+
{d\over d \ln R}\left({S_3(R) \sigma_R \over 6} \right) H_2(\delta_v / \sigma_R)
+{1 \over 2}{d\over d\ln R}\left({S_3(R) \sigma_R \over 6} \right)^2 H_5(\delta_v / \sigma_R) 
+{1 \over 6}{d\over d\ln R}\left({S_3(R) \sigma_R \over 6} \right)^3 H_8(\delta_v / \sigma_R) \cr\cr
&& +
{d\over d\ln R}\left({S_4(R) \sigma_R^2 \over 24} \right) H_3(\delta_v / \sigma_R)
+{1 \over 2}{d\over d\ln R}\left({S_4(R) \sigma_R^2 \over 24} \right)^2 H_7(\delta_v / \sigma_R) 
+{1 \over 6}{d\over d\ln R}\left({S_4(R) \sigma_R^2 \over 24} \right)^3 H_{11}(\delta_v / \sigma_R)~.
\Biggr\}
\end{eqnarray}
Following the previous section, we define the ratio between the void abundance with the pure Gaussian PDF
and that with the primordial non-Gaussian corrections as
\begin{eqnarray}
{\cal R}_{\rm NG}^{\rm void} \equiv {dn^{\rm void}(R)/dR \over dn^{\rm void}_{\rm G}(R) / dR}~. \nonumber\\
\label{eq:voidratio}
\end{eqnarray}
In Fig.~\ref{fig:void_ratio.eps}, 
we show ${\cal R}_{\rm NG}^{\rm void}$ for the cases with $f_{\rm NL}=100, \tau_{\rm NL} = g_{\rm NL} = 0$ (red lines),
$\tau_{\rm NL} = 10^6, f_{\rm NL} = g_{\rm NL} = 0$ (blue thick lines)
and $\tau_{\rm NL} = 10^4, f_{\rm NL} = g_{\rm NL} = 0$ (black thin lines).
We adopt $\delta_v = -0.7$.
We also show the ratio with changing the redshift; the solid lines for the case with $z=0.0$,
the dot-dashed lines for the case with $z=0.75$ and the dashed lines for the case with $z=1.5$.
From this figure,
we conclude that
the non-Gaussian void abundance
with  non-zero $\tau_{\rm NL}$
becomes larger than the Gaussian one
on relatively larger scales
whereas
that with the non-zero $f_{\rm NL}$
becomes smaller.
On the other hand, as seen in the previous section,
the halo abundance becomes larger not only with non-zero $\tau_{\rm NL}$
but also with  non-zero $f_{\rm NL}$ in relatively more massive objects.
Hence, from this discussion, we confirm that
the non-Gaussian effects on both the halo and the void abundances allow as to
distinguish the large kurtosis, i.e., large $\tau_{\rm NL}$, case
from the large skewness, i.e., large $f_{\rm NL}$ case~\citep{Chongchitnan:2010xz}.

\section{Summary and Discussion}
\label{sec:sum}

It has recently become clear that
cosmological large-scale structure and CMB observations
could provide stringent constraints
on the PDF of primordial adiabatic curvature fluctuations.
In particular,
the high order moments of the PDF, such as
its skewness and kurtosis,
can give unique insights into the dynamics and conditions of the
inflationary phase in the early Universe.

In this paper, we have investigated the effects of  the $\tau_{\rm NL}$-type of primordial
non-Gaussianity on the halo mass function.
In particular,
we have obtained a formula for the halo mass function with the non-Gaussian
corrections coming from the kurtosis induced by
the non-zero $\tau_{\rm NL}$.
We find that
the deviations of the non-Gaussian mass function from the Gaussian one
become larger for  larger mass objects 
($M \gtrsim10^{14} h^{-1} M_\odot$ for $z \sim 0$)
as well as at higher redshifts ($z \gtrsim 1$ for $M \approx 10^{14} h^{-1} M_\odot$)
in the case with $\tau_{\rm NL} = O(10^6)$.
Such features are quite similar to those
obtained from skewness-driven
non-Gaussian corrections
that are induced by the $f_{\rm NL}$-type of primordial non-Gaussianity.

As examples of applications of our formulae,
we have considered
the effects on early star formation, formation  of the most
massive objects at  high redshift,
and the abundance of voids.

For early star formation,
we applied our formula for the non-Gaussian halo mass function
in order to estimate of the redshift-dependence of the cumulative number
of photons emitted from population III stars,
a crucial quantity
in considerations of the reionization history of the Universe.
We found that primordial non-Gaussianity 
does not affect the reionization history of the Universe
on the average,
but at high redshift ($z\simeq 20$), namely
the earliest stages of  reionization,
it is effective.

We have also obtained an estimate of the value of $\tau_{\rm NL}$
needed to naturally explain the existence of the  galaxy cluster  XMMU J2235.3-2557,
namely  $\tau_{\rm NL} = 1.7 \times 10^6 $.
Hence,
in light of the result of Smidt et al.,
we might need to consider a possibility such as  scale-dependent $\tau_{\rm NL}$
in the case with non-zero $f_{\rm NL}$.  
In Ref.~\citep{Hoyle:2010ce},
the authors have investigated $15$ high-mass and high-redshift galaxy clusters
and found that such objects are extremely rare in the standard $\Lambda$CDM
model with  Gaussian primordial fluctuations.
They derived a
constraint on $f_{\rm NL}$ in order to
explain the mere existence of these objects as $f_{\rm NL} > 475$ at $95\%$ confidence
level,  with the other cosmological parameters fixed to best fit values
of WMAP data.
In Ref,~\citep{Enqvist:2010bg},
the authors have extended the analysis of Ref.~\citep{Hoyle:2010ce} to the case
with non-zero $g_{\rm NL}$.
It should clearly be of interest  to derive a constraint on $\tau_{\rm NL}$
for these observed high-mass and high-redshift galaxy clusters.
We will address this in future work.

As mentioned in Refs.~\citep{Kamionkowski:2008sr,Chongchitnan:2010xz},
the non-Gaussian correction coming from  skewness
reduces the abundance of voids on large scales when the non-linearity
parameter $f_{\rm NL}$ is positive in contrast to the fact that
positive $f_{\rm NL}$ enhances the number of more massive halo objects.
On the other hand,
the non-Gaussian correction coming from kurtosis
enhances not only the numbers of more massive halo objects
but also the abundances of voids on large scales.
Hence, if one could also measure the void abundance as well as  the
halo mass function more precisely,
one could potentially distinguish between the $f_{\rm NL}$ 
and the $\tau_{\rm NL}$-types of  primordial non-Gaussianity.

\textit{
NOTE;
}
~During the time that we were preparing this manuscript,
Ref.~\citep{LoVerde:2011iz} appeared on the arXiv.
In Ref.~\citep{LoVerde:2011iz},
they  considered the same type of primordial non-Gaussianity
as  in our study and
obtained a useful analytic formula for the halo mass function
with the kurtosis type primordial non-Gaussianity using N-body simulations.
We find that 
our formula~(\ref{eq:nonGaussmass}) is in reasonably good agreement with 
their formula as far as the behavior of the halo mass function
with the kurtosis type of  primordial non-Gaussianity.

\section{acknowledgments}

S.Y. thanks Yoshitaka Takeuchi and Shogo Masaki, and J.S. thanks Laura Cayon, Sirichai Chongchitnan, and Christopher Gordon for 
valuable discussions.
This work is supported by the
Grant-in-Aid for Scientific research from the Ministry of Education,
Science, Sports, and Culture, Japan, No. 22340056. 
The authors also acknowledge support from 
the Grant-in-Aid for Scientific Research
on Priority Areas No. 467 ``Probing the Dark Energy through an
Extremely Wide and Deep Survey with Subaru Telescope'' and
the Grant-in-Aid for
the Global COE Program ``Quest for Fundamental Principles in the
Universe: from Particles to the Solar System and the Cosmos'' from
MEXT, Japan.
This work is also supported in part by World Premier International
Research Center Initiative, MEXT, Japan.

\appendix
\section{MVJ expression}
\label{app:mvj}

In Ref.~\citep{Matarrese:2000iz}, the authors have given a formula
for the ratio between the non-Gaussian mass function and the Gaussian mass function
as
\begin{eqnarray}
R_{\rm NG}^{\rm MVJ} (M,z)
&=& \exp \left[\nu_c^3 {S_3(R) \sigma_R \over 6} + \nu_c^4 {S_4(R) \sigma_R^2 \over 24} \right]
 \times
\left[ \delta_3 + {\nu_c \over \delta_3}
\left( - {S_3(R) \sigma_R \over 6} \right) 
+ \left({ d \ln \sigma_R \over dM} \right)^{-1}
{d \over dM}\left( {S_3(R) \sigma_R \over 6} \right)\right]
\cr\cr
&& \qquad\qquad \times
\left[ \delta_4 + {\nu_c^2 \over \delta_4}
\left( - {S_4(R) \sigma_R^2 \over 12} \right) 
+ \left({ d \ln \sigma_R \over dM} \right)^{-1}
{d \over dM}\left( {S_4(R) \sigma_R^2 \over 24} \right)\right]~, \cr\cr
\delta_3
& \equiv &
\left( 1 - \nu_c {S_3(R) \sigma_R \over 3} \right)^{1/2}~,~
\delta_4
 \equiv
\left( 1 - \nu_c^2 {S_4(R) \sigma_R^2 \over 12} \right)^{1/2}~,
\end{eqnarray}
which was not derived based on the Edgeworth expansion as mentioned in section~\ref{sec:nGmass}.
In section~\ref{sec:nGappl}, we have discussed some applications of the non-Gaussian halo mass function.

As for the discussion in subsection~\ref{sub:early} about early star formation,
the redshift-dependence of
the critical value of the cumulative photon number per H atom ($n_\gamma / n_H = 10$)
is not so sensitive to primordial non-Gaussianity.
In order to estimate more precisely how $n_\gamma / n_H$ at high redshift is enhanced 
due to primordial non-Gaussianity,
we should check which formula better describes the effect of primordial non-Gaussianity
on the halo mass function. This is a future issue.
In subsection~\ref{sub:void}, we have discussed the void abundance
and noted that the kurtosis type of primordial non-Gaussianity
can enhance the abundance of the large voids
as opposed to the skewness type of primordial non-Gaussianity.
This is just qualitative discussion.

On the other hand, the discussion in subsection~\ref{sub:mass} is so quantitative
and hence we have investigated the difference of
the estimated value of $\tau_{\rm NL}$ for the observation of XMMU J2235.3-2557
between the case with MVJ expression and that with Eq.~(\ref{eq:nonGaussmass})
given in section~\ref{sec:nGmass}.
Our naive estimated value of $\tau_{\rm NL}$ given in subsection~\ref{sub:mass}
is $\tau_{\rm NL} = 1.7 \times 10^6$.
For the case by making use of MVJ expression, we obtained $\tau_{\rm NL} = 1.1 \times 10^6$.
These values seem to be same order and hence the result does not extremely change.


\begin{thebibliography}{}




\bibitem[Barkana \& Loeb (2001)]{Barkana:2000fd}
  Barkana~R.,~Loeb~A.,
  2001,
  Phys. Rep.,  349, 125



\bibitem[Bartolo et al. (2004)]{Bartolo:2004if}
  Bartolo~N.,~Komatsu~E.,~Matarrese~S.,~Riotto~A., 2004
  Phys. Rep., 402, 103


\bibitem[Bartolo, Matarrese \& Riotto (2010)]{Bartolo:2010qu}
  Bartolo~N.,~Matarrese~S. and~Riotto A., 2010,  
  Advances in Astronomy, 2010, 157079


\bibitem[Bryan \& Norman (1998)]{Bryan:1997dn}
  Bryan~G.~L.,~Norman M.~L., 1998,
  ApJ, 495, 80


\bibitem[Byrnes \& Choi (2010)]{Byrnes:2010em}
  Byrnes C.~T.,~Choi K.~Y., 2010,
  Advances in Astronomy, 2010, 724525






\bibitem[Cayon, Gordon \& Silk(2010)]{Cayon:2010mq}
  Cayon L., Gordon C.,~Silk J., 2010, preprint
  (arXiv:1006.1950)
  
\bibitem[Chongchitnan \& Silk(2010a)]{Chongchitnan:2010xz}
  Chongchitnan S.,~Silk J., 2010,
  ApJ, 724, 285
  
\bibitem[Chongchitnan \& Silk(2010b)]{Chongchitnan:2010hb}
  Chongchitnan S.,~Silk J., 2011,
  Phys.\ Rev.\  D, 83, 083504
\bibitem[Colberg et al.(2008)]{Colberg:2008qg}
  Colberg J.~M.~et al., 2008, MNRAS, 387 933


\bibitem[D'Amico et al.(2010)]{D'Amico:2010ta}
  D'Amico G., Musso~M., Norena~J., Paranjape A., 2011, 
  J. Cosmology Astropart. Phys., 02, 01
 
 
\bibitem[De Simone, Maggiore \& Riotto(2010)]{DeSimone:2010mu}
  De Simone~A., Maggiore M., Riotto A., 2010, preprint
  (arXiv:1007.1903)


\bibitem[Desjacques \& Seljak(2010)]{Desjacques:2009jb}
 Desjacques V.,~Seljak U., 2010,
  Phys.\ Rev.\  D, 81, 023006



\bibitem[Enqvist \& Sloth(2002)]{Enqvist:2001zp}
Enqvist~K.,~S.~Sloth~M., 2002,
Nucl.\ Phys.\ B, 626, 395


\bibitem[Enqvist, Hotchkiss \& Taanila(2010)]{Enqvist:2010bg}
  Enqvist~K., Hotchkiss~S., Taanila~O., 2010, preprint
  (arXiv:1012.2732)




\bibitem[Fergusson, Regan \& Shellard(2010)]{Fergusson:2010gn}
  Fergusson J. R., Regan D. M., Shellard E. P. S., 2010, preprint
  (arXiv:1012.6039).


\bibitem[Fuller \& Couchman(2000)]{Fuller:2000vk}
  Fuller T. M., Couchman H. M. P., 2000,
  ApJ, 544, 6


\bibitem[Gangui et al.(1994)]{Gangui:1993tt}
  Gangui A., Lucchin F., Matarrese S., Mollerach S.,
  1994, ApJ, 430, 447

\bibitem[Grossi et al.(2009)]{Grossi:2009an}
  Grossi M., Verde L., Carbone C., Dolag K., Branchini E.,
  Iannuzzi F., Matarrese S., Moscardini L., 2009
  MNRAS, 398, 321



\bibitem[Hoyle, Jimenez \& Verde(2010)]{Hoyle:2010ce}
  Hoyle B., Jimenez R., Verde L., 2011,
  Phys.\ Rev.\ D, 83, 103502

\bibitem[Huang(2009)]{Huang:2009vk}
  Huang Q.~G., 2009,
  J. Cosmology Astropart. Phys., 06, 35



\bibitem[Ichikawa et al.(2008)]{Ichikawa:2008iq}
  Ichikawa K., Suyama T., Takahashi T., Yamaguchi M., 2008,
  Phys.\ Rev.\  D, 78, 023513




\bibitem[Jee et al.(2009)]{Jee:2009nr}
  Jee M. J. et al., 2009,
  ApJ,  704, 672


\bibitem[Juszkiewcz et al.(1995)]{Juszkiewicz:1993hm}
  Juszkiewicz R., Weinberg D. H., Amsterdamski P., Chodorowski M., Bouchet F.,
  1995, ApJ,  442, 39




\bibitem[Kamionkowski, Verde \& Jimenez(2009)]{Kamionkowski:2008sr}
  Kamionkowski M., Verde L., Jimenez R., 2009,
  J. Cosmology Astropart. Phys., 01, 10


\bibitem[Komatsu \& Spergel(2001)]{Komatsu:2001rj}
  Komatsu E., Spergel D. N., 2001,
  Phys.\ Rev.\  D, 63, 063002


\bibitem[Komatsu et al.(2011)]{Komatsu:2010fb}
  Komatsu E. et al.  [WMAP Collaboration], 2011,
  ApJ, Suppl.,  192, 18


\bibitem[Komatsu(2010)]{Komatsu:2010hc}
  Komatsu E., 2010,
  Classical and  Quantum Gravity,  27, 124010


\bibitem[Langlois \& Vernizzi(2004)]{Langlois:2004nn}
  Langlois D., Vernizzi F., 2004,
  Phys. Rev.  D, 70, 063522

  
\bibitem[LoVerde et al.(2008)]{LoVerde:2007ri}
  LoVerde M., Miller A., Shandera S., Verde L., 2008,
  J. Cosmology Astropart. Phys., 04, 14



\bibitem[LoVerde \& Smith(2011)]{LoVerde:2011iz}
  LoVerde M., Smith K. M., 2011, preprint
   (arXiv:1102.1439)


\bibitem[Lyth \& Wands(2002)]{Lyth:2001nq}
Lyth D. H., Wands D., 2002,
Phys.\ Lett.\ B,  524, 5 




\bibitem[Matarrese, Verde \& Jimenez(2000)]{Matarrese:2000iz}
  Matarrese S., Verde L., Jimenez R., 2000,
  ApJ, 541, 10

\bibitem[Maggiore \& Riotto(2010)a]{Maggiore:2009rx}
  Maggiore M., Riotto A., 2010,
  ApJ,  717, 526
  
\bibitem[Maggiore \& Riotto(2010)b]{Maggiore:2009hp}
  Maggiore M., Riotto A., 2010,
  MNRAS, 405, L1244
  


\bibitem[Moroi \& Takahashi(2001)]{Moroi:2001ct}
Moroi T., Takahashi T., 2001,
Phys.\ Lett.\ B, 522, 215
[Erratum-ibid.\ B, 539, 303]





\bibitem[Park \& Lee(2007)]{Park:2006wu}
  Park D., Lee J., 2007,
  Phys.\ Rev.\ Lett., 98, 081301




\bibitem[Rosati et al.(2009)]{Rosati:2009cm}
  Rosati P. et al., 2009,
  A \& A, 508, 583


\bibitem[Salopek \& Bond(1990)]{Salopek:1990jq}
  Salopek D. S., Bond J. R., 1990,
  Phys.\ Rev.\  D, 42, 3936

\bibitem[Shandarin et al.(2005)]{Shandarin:2005ea}
  Shandarin S., Feldman H. A., Heitmann K., Habib S., 2006
  MNRAS, 367, 1629


 
\bibitem[Slosar et al.(2008)]{Slosar:2008hx}
  Slosar A., Hirata C., Seljak U., Ho S., Padmanabhan N., 2008,
  J. Cosmology Astropart. Phys., 08, 31

\bibitem[Smidt et al.(2010)]{Smidt:2010ra}
  Smidt J., Amblard A., Byrnes C. T., Cooray A., Heavens A., Munshi~D., 2010,
  Phys.\ Rev.\  D, 81, 123007 (2010)


\bibitem[Smith \& LoVerde(2010)]{Smith:2010gx}
  Smith K. M., LoVerde M., 2010, preprint
   (arXiv:1010.0055)

\bibitem[Somerville \& Livio(2003)]{Somerville:2003sk}
  Somerville R. S., Livio M., 2003,
  ApJ, 593, 611

\bibitem[Somerville, Bullock \& Livio(2003)]{Somerville:2003sh}
  Somerville R. S., Bullock J. S., Livio M.,
  2003, ApJ, 593, 616


\bibitem[Sugiyama, Zaroubi \& Silk(2004)]{Sugiyama:2003tc}
  Sugiyama N., Zaroubi S., Silk J.,
  2004, MNRAS, 354, 543


 
\bibitem[Sugiyama, Komatsu \& Futamase(2011)]{Sugiyama:2011jt}
  Sugiyama N. S., Komatsu E., Futamase T., 2011, preprint
   (arXiv:1101.3636)



\bibitem[Suyama \& Yamaguchi(2008)]{Suyama:2007bg}
  Suyama T., Yamaguchi M., 2008,
  Phys.\ Rev.\  D, 77, 023505
  
\bibitem[Suyama et al.(2010)]{Suyama:2010uj}
  Suyama T., Takahashi T., Yamaguchi M., Yokoyama S.,
  2010,
  J. Cosmology Astropart. Phys. 12, 30
   




\bibitem[Tseliakhovich, Hirata \& Slosar(2010)]{Tseliakhovich:2010kf}
  Tseliakhovich D., Hirata C., Slosar A.,
  2010,
  Phys.\ Rev.\ D, 82, 043531



\bibitem[Verde et al.(2000)]{Verde:1999ij}
  Verde L., Wang L. M., Heavens A., Kamionkowski M.,
  2000, MNRAS, 313, L141

\bibitem[Verde et al.(2001)]{Verde:2000vr}
  Verde L., Jimenez R., Kamionkowski M., Matarrese S.,
  2001,
  MNRAS, 325, 412


\bibitem[Verde(2010)]{Verde:2010wp}
  Verde L., 2010,
  Advances in Astronomy, 2010, 768675


\bibitem[Wagner, Verde \& Boubekeur(2010)]{Wagner:2010me}
  Wagner C., Verde L., Boubekeur L.,
  2010,
  J. Cosmology Astropart. Phys. 10, 22



\bibitem[Yoshida et al.(2003)]{Yoshida:2003rw}
  Yoshida N., Abel T., Hernquist L., N.~Sugiyama N.,
  2003,
  ApJ,  592, 645







\end{thebibliography}
\end{document}